\documentclass[twocolumn,aps]{revtex4}
\usepackage{graphicx,amsmath,epsfig}
\usepackage{indentfirst,here}

\begin{document}

\title{Multiscale Analysis of the Stress State in a Granular Slope in Transition to Failure}

\author{Lydie Staron$^1$, Jean-Pierre Vilotte$^2$, and Farhang Radjai$^3$}
\affiliation{$1$ Department of Applied Mathematics and Theoretical Physics,
University of Cambridge, CB3 0WA Cambridge, UK\\
$^2$ Institut de Physique du Globe de Paris, F-75252 Paris cedex 05, France. \\
$^3$ LMGC, CNRS-Universit\'e
Montpellier II, F-34095 Montpellier cedex, France.}

\begin{abstract}
By means of contact dynamics simulations, we analyze the stress state in a granular bed slowly tilted towards its angle of repose. An increasingly large number of grains are overloaded in the sense that they are found to carry a stress ratio above the Coulomb yield threshold of the whole packing. Using this property, we introduce a coarse-graining length scale at which all stress ratios are below the packing yield threshold.
We show that this length increases with the slope angle and jumps to a length comparable to the depth of the granular bed at an angle below the angle of repose. This transition coincides with the onset of dilatation in the packing. We map this transition into a percolation transition of the overloaded grains, and we argue that in the presence of long-range correlations above the transition angle, the granular slope is metastable.
\end{abstract}

\date{\today}


\maketitle

\section{Introduction}
\label{intro}

The science of granular materials was initiated by Coulomb's analysis of the equilibrium and failure of a granular talus~\cite{coulomb73}. The well-known Coulomb's failure criterion was later incorporated in the framework of a rigid-plastic behaviour based on experimental testing of granular samples with homogeneous boundary conditions~\cite{drucker54,wood_book,oda99}. Two centuries after Coulomb, the slope failure phenomena continue to interest scientists from various fields with evident applications to geological processes and industrial handling of granular materials~\cite{jaeger_nagel96,degennes99}. The main reason is that the phenomena involved in the evolution of a granular slope are richer than what might be expected from a mean macroscopic analysis~\cite{jaeger89,darve00,daerr00}. On the other hand, the mechanisms that lead to slope failure have not yet been well understood from a grain scale standpoint~\cite{quartier00,alonso98,lemaitre02,staron02}.

New investigation tools, such as fine imaging techniques and discrete numerical simulations, have shown that granular media are very inhomogeneous at the grain scale, and the microstructure, {\it i.e.} the organization of the grains and their contacts in space, can evolve in many different ways in response to external loading and boundary conditions~\cite{oda72a,oda80,rothenburg89,cambou93,radjai98,radjai01a}. Often, large fluctuations are observed in the course of shearing and from one situation to another~\cite{howell99}. Several observations suggest that surface failure may occur at slope angles well below the angle of repose~\cite{darve99}, and metastable states exist in the vicinity of the angle of repose~\cite{daerr00}.

In this context, a closer look at the microstructure and spatio-temporal scales governing the behaviour of a granular slope can not be avoided. The query is which internal variables or order parameters represent the evolution of a granular slope towards surface failure~\cite{roux01}. Even under ``quasi-static'' conditions, the grains in a cohesionless granular medium exhibit a high degree of mobility. Both dynamical instabilities and collective rearrangement phenomena occur frequently in response to slightest load increments~\cite{moreau97,radjai02,staron02}. This observation shows that, even in a granular medium far from macroscopic failure, the failure conditions are often fulfilled locally. Such effects may be observed by looking at grain displacements or contact forces at different scales.

In this paper, we focus on the scaling of local stresses in a two-dimensional granular bed simulated by the Contact Dynamics method. The bed is tilted slowly towards its angle of repose $\theta_c$ at which slope failure occurs. This slow gravity loading gives rise to grain rearrangements that harden the bed. Without such a hardening (or plastification) process, the bed can not reach its angle of repose. For example, a bed prepared initially by pourring the grains onto a rough horizontal plane, when tilted suddenly to a finite slope $\theta$, will fail immediately. Hence, the evolution of local stress states as a function of loading is important for understanding how the failure limit is reached and in which respects it is controlled by the details of the microstructure.

For the analysis of stresses at different scales, we need a quantity that plays the same role as the Cauchy stress tensor in continuous media. It can be shown that the concept of ``internal moment tensor'', introduced by Moreau~\cite{moreau97}, generalizes consistently the Cauchy stress tensor to discrete media and, what is more, its mechanical content with respect to Newton's equations of motion remains the same whether applied to a single grain or to a collection of grains inside a control volume.

In the following, we first introduce the numerical procedures and the concept of stress tensor in terms of internal moments which will be used throughout the paper. Then, we analyze the data from a global standpoint, {\it i.e.} by considering the evolution of average stresses in the granular bed as a whole. We discuss also a theoretical evaluation of the stresses in the presence of walls in comparison to a bed with periodic boundary conditions. Then, we analyze local and coarse-grained stresses as a function of the tilt angle. We will conclude with a summary of our main results and a discussion about their physical interpretation.

%--------------------------------------------------------------------------
\section{Numerical procedures}
\label{expenum}

\subsection{Simulation method}
For a discrete simulation of the dynamics of a collection of cohesionless rigid grains, two strategies can be adopted, differring mainly in the implementation of contact repulsion and friction. In the popular Molecular Dynamics (MD) method a repulsive potential is introduced as a function of contact interpenetration and the friction force obeys an elastic or viscous law up to a Coulomb threshold~\cite{cundall79,herrmann92,luding98c}. The equations of motion are explicitely integrated according to different classical schemes. As an interesting alternative to the MD method, the Contact Dynamics (CD) approach deals directly with infinitely stiff contact laws or, more generally ``nonsmooth'' laws~\cite{moreau88,jean92,jean_moreau92,moreau94,jean95,radjai98}.
Hence, in contrast to the MD method, no elastic stiffness or viscous regularization of the Coulomb friction law are to be introduced at the contact level. This allows for larger time steps and thus for a substential reduction of computational time.
At the same time, no purely computational parameters are introduced, the only parameters being the intergrain coefficient of friction $\mu$ and the normal and tangential coefficients of restitution that account for dissipation in the advent of collisions between the grains. In application to quasi-static deformations of a granular packing, one expects similar behaviours from both methods as far as large time scales (beyond the elastic response time) and plastic deformations are concerned. In the present work, we applied the CD method.

\subsection{System characteristics}
\label{system}

We consider circular grains in two dimensions, with diameters uniformly distributed in the interval $[D_{min},D_{max}]$ with $D_{max}/D_{min} = 1.5$.
This slight polydispersity reduces long-range cristal-like ordering in the packing. The grains interact only through frictional cohesionless contacts with a coefficient of friction $\mu = 0.5$. The same value is used for grain-wall contacts. Moreover, we assume zero coefficients of restitution between grains. Slightly larger coefficients of restitution do not influence the mechanical behaviour in a dense granular packing where multiple contacts dissipate efficiently the kinetic energy.

The granular beds were prepared by random deposition of grains on a horizontal plane in the gravity field. The plane was made rough by sticking grains of diameter $D$, $D$ being the mean diameter of the grains in the packing. Two different boundary conditions were implemented: wall boundary conditions (WBC) where the bed is confined between two vertical walls, and periodic  boundary conditions (PBC) in the horizontal direction. We prepared different beds having all the same depth $H \simeq 40D$, but different lengths $L$. In the following, we will analyze WBC systems with $L = 100D, 150D, 200D, 250D$ and $300D$, corresponding to 4000, 6000, 8000, 10000 and 12000 grains respectively, as well as a PBC system composed of 8000 grains with $L=200$, where $L$ refers in that case to the length of the simulation cell.

The general features of the packings are the same independently of the boundary conditions. They have a solid fraction $\rho \simeq 0.8$ and a coordination number $z \simeq 3.5$, corresponding to a random close packing. Despite the polydispersity introduced in the grain size distribution, we observe privileged directions of contact normals at $0^\circ$, $60^\circ$ and $120^\circ$ with respect to the horizontal direction. A signature of this ordering appears in the distribution of local stresses, as we shall see below.

\begin{figure}
\centerline{\epsfig{file = 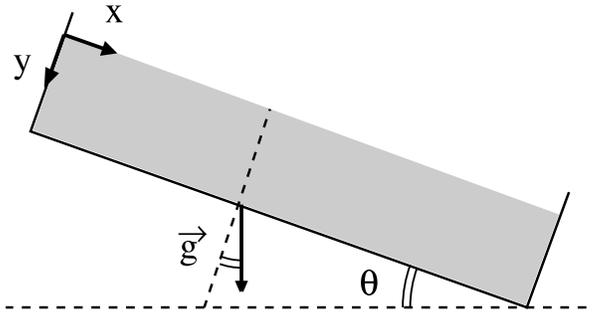,angle = 0,width = 0.95\linewidth}}
\caption{Schema of the simulation: a granular bed with wall boundary conditions is tilted in the gravity field to bring the slope of the free surface $\theta$ from originally $0$ up to the angle of avalanche $\theta_c$.}
\label{Fig1}
\end{figure}

The granular beds are tilted at a constant rotation rate $\omega = 1^\circ s^{-1}$ in the gravity field. The slope of the free surface $\theta$ increases monotonically from $\theta = 0$ to the angle of repose $\theta = \theta_c$ (Figure~\ref{Fig1}). At this point, the stability limit of the packing is reached and a surface flow is triggered. Analyzing a set of 50 independent simulations, we checked that both the average and local behaviours that will be discussed in this paper are highly reproducible.

%-----------------------------------------------------
\section{The stress tensor}
\label{definitions}

The dynamics of a granular system is naturally described in terms of grain degrees of freedom (velocities) and contact actions, including normal and tangential forces as well as contact torques (in a cohesive granular medium).
This vectorial description is, however, unsuitable in a macroscopic formulation of the rheological behavior where the material has to be described as an effective continuous medium where the stress and strain variables are stress tensors $\boldsymbol \sigma$ and strain tensors $\boldsymbol \varepsilon$ defined over representative {\em volumes} (in contrast to forces and velocities which act over {\em points}).

\begin{figure}
\centerline{\epsfig{file = 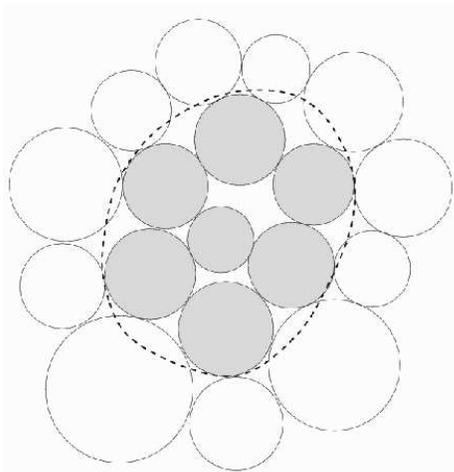,angle = 0,width = 0.7\linewidth}}
\caption{The stress tensor in the volume bounded by the dashed line can be evaluated as the surface density of the forces exerted by the external grains on the grains located inside the volume.}
\label{Fig2}
\end{figure}

In principle, it is not difficult to evaluate the stress components in a control volume by simply calculating the surface density of forces applied by the grains situated at one side of a surface (a line in 2D) to the grains situated on the other side (Figure~\ref{Fig2}). In this sense, the Cauchy stress tensor is as well defined as in other materials (with or without a granular structure). In the case of dilute granular materials, the convection of grain momenta is the main mechanism of stress transmission as in a classical gaz. Here, we are concerned only with quasi-static deformations of a granular material for which contact actions play the main role in stress transmission.
Using the above operational definition, it is possible to extract an expression for the stress components $\sigma_{ij}$ in terms of contact forces $\{ {\boldsymbol f}^\alpha \}$ and the vectors ${\boldsymbol \ell}^\alpha$ joining the particle centers:
\begin{equation}
\sigma_{ij} = \frac{1}{V} \sum_{\alpha \in V} f_i^{\alpha} \ell_j^\alpha,
\label{eqn1}
\end{equation}
where $V$ is the control volume and $\alpha$ denotes the contacts in V.

Several authors have proposed different methods to demonstrate this or similar relations~\cite{christoffersen81,kruyt96,goddard98}. Such expressions are well-defined if evaluated over a large volume containing many contacts.

There is, of course, no reason for not using the expression~\ref{eqn1} at smaller scales down to a single grain if the volume $V$ is properly adjusted. However, strictly speaking, the result will not be a stress tensor in the operational sense defined above. This means that the problem should be posed in reverse order: Is there a quantity whose physical content remains the same whether applied to a single grain or to a collection of grains inside a control volume and that tends to the Cauchy stress tensor at large scales? Moreau showed that the concept of {\em internal moment tensor} fulfills these conditions~\cite{moreau97}.
For clarity, we briefly introduce this concept below.

In the framework of the virtual power formalism, a force (in the general sense) experienced by a bounded portion $S$ of a material system is defined through the expression of the power ${\cal P}$ that it develops when subjected to a virtual velocity field $v({\boldsymbol r})$.
Let $v({\boldsymbol r})$ be an affine field,
\begin{equation}
v_i ({\boldsymbol r})= v_i(0) + b_{ij} r_j,
\label{eqn2}
\end{equation}
where we assume Eintein's summation rule over subscripts. By definition, the power ${\cal P}_{int}({\boldsymbol v})$ of internal forces is linear in ${\boldsymbol v}$.
This means that there exist $\boldsymbol R$ and $\boldsymbol M$ such that
\begin{equation}
{\cal P}_{int} = R_i v_i(0)+M_{ij}b_{ij}
\label{eqn3}
\end{equation}
In the particular case of a rigid body motion, $\boldsymbol b$ is antisymmetric ($b_{ij} = -b_{ji}$) and ${\cal P}_{int}=0$ by virtue of Newton's third law. This implies that ${\boldsymbol R} =0$ and ${\boldsymbol M}$ is a symmetric tensor of rank 2 and independent of the choice of the reference frame. Following Moreau, we will refer to ${\boldsymbol M}$ as the internal moment tensor of the system~\cite{moreau97}.

\begin{figure}
\centerline{\epsfig{file = 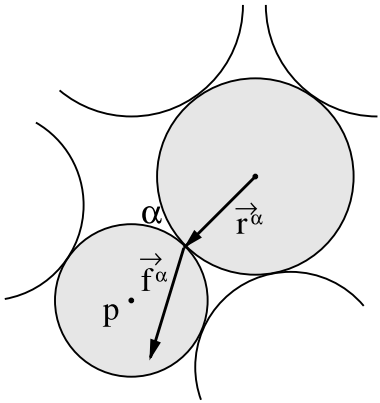,angle = 0,width = 0.6\linewidth}}
\caption{Force ${\boldsymbol f}^\alpha$ applied on a grain $p$ at the contact point ${\boldsymbol r}^\alpha$ by a neighbouring grain.}
\label{Fig3}
\end{figure}

By definition, the internal moment tensor makes sense at all scales. In particular, we may evaluate the internal moment tensor of a grain within a granular system.
The simplest example occurs when the system is in static equilibrium.
In this case, the total power ${\cal P} = {\cal P}_{int}+{\cal P}_{ext}$, where ${\cal P}_{ext}$ is the power associated with external forces, is zero independently of the choice of the virtual velocities. If the only forces ${\boldsymbol f}^\alpha$ acting on a grain $p$ are those exerted at its contact points $ {\boldsymbol r}^\alpha$ by neighbouring grains, then the internal power is ${\cal P}_{int}(p)= -{\cal P}_{ext}(p)=-\sum_{\alpha \in p} v_i ({\boldsymbol r}^\alpha) f_i^\alpha$ (Figure~\ref{Fig3}).
Identifying this with the general expression~\ref{eqn3} of the internal power, we get $M_{ij}(p)  = -\sum_{\alpha \in p} r_i^\alpha f_j^\alpha$.
If the condition of equilibrium does not apply, the total virtual power is given by $\int \gamma({\boldsymbol r}) dm$, where $\gamma({\boldsymbol r})$ is the acceleration field and $dm$ denotes the mass measure.

In the case of circular grains with moment of inertia $I$ about the grains centers, the general expression of the internal moment tensor of a grain $p$ becomes~\cite{moreau97}
\begin{equation}
M_{ij}(p)  = - \sum_{\alpha \in p} r_i^\alpha f_j^\alpha - \frac{1}{2}I
\omega^2 \delta_{ij},
\label{eqn4}
\end{equation}
where $\omega$ is the rotation velocity and $\delta_{ij}$ is the Kronecker symbol. It can be shown that the expression~\ref{eqn4} holds also in the presence of bulk forces (gravity) acting at grain centers if the origin of coordinates for each grain is placed at its center.

The internal moment ${\boldsymbol M}(p_1 \cup p_2)$ of two grains $p_1$ and $p_2$ sharing a contact is the sum of their respective internal moments ${\boldsymbol M}(p_1)$ and ${\bm M}(p_2)$ because opposite reaction forces of equal magnitude act on the two grains at the {\em same} contact point.
This additive property implies that the total internal moment ${\boldsymbol M}(S)$ of a system $S$ is simply the sum of the internal moments of all grains included in $S$. On the other hand, if the number of grains in $S$ is sufficiently large, it makes sense to evaluate the Cauchy stress tensor ${\boldsymbol \sigma}$ for $S$. Assuming the same test field as~\ref{eqn2}, the corresponding internal power by definition of ${\boldsymbol \sigma}$ is
\begin{equation}
{\cal P}_{int}=\int_V \sigma_{ij} \partial_i v_j dV.
\label{eqn5}
\end{equation}
Then, according to~\ref{eqn3}, we have
\begin{equation}
M_{ij} (S) = \int_V \sigma_{ij} dV = \langle \sigma_{ij} \rangle V.
\label{eqn6}
\end{equation}
This shows that the internal moment tensor of $S$ per unit volume (${\boldsymbol M}/V$) tends to the average Cauchy stress tensor $\langle \sigma_{ij} \rangle$ at larger scales or for an increasing number of grains contained in $S$.

We see that the internal moment tensor has all the required properties for a scaling analysis of stress transmission in granular media. Conceptually, the internal moment tensor per unit volume in a discrete system plays the same role as the Cauchy stress tensor in a continuous medium. For this reason, we will refer to the internal moments of the grains as {\em grain stresses}. In the following, we will be more specifically interested in stress ratios, {\it i.e.} the ratio of the deviatoric part of grain stresses normalized by the corresponding spherical part.

%------------------------------------------------------------
\section{The Packing stress}
\label{macrostress}

In this section, we study the average (or total) stress of the granular bed as a function of the tilt angle $\theta$. In the absence of plastification, {\it i.e.} for a rigid-plastic behaviour without  rearrangements, the average stress tensor ${\boldsymbol \sigma}$ follows directly the rotation of the bed with respect to the direction of gravity (and the volume of the bed remains constant)~\cite{roux01}.
Here, we check whether the simulations yield a picture close to this prediction in spite of rearrangement phenomena as the bed is tilted.
The packing stress tensor is evaluated for the whole packing by simply adding up the grain stresses and dividing by the total volume of the packing.
We consider more specifically the ratio of the normal component $\sigma_N$ of the stress tensor along the direction of the free surface to the tangential component $\sigma_T$.
Alternatively, we evaluate the stress ratio $\Gamma$ defined as the stress deviator $Q=\sigma_1 - \sigma_2$, where $\sigma_1$ and $\sigma_2$ are the principal values, normalized by the average stress $P=\sigma_1 + \sigma_2$:
\begin{equation}
\Gamma = \frac{Q}{P}.
\label{Gamma}
\end{equation}
The Coulomb criterion implies $\Gamma=\sin \theta_c$, where $\theta_c$ is the angle of repose, at incipient failure.

\begin{figure}
\centerline{\epsfig{file = 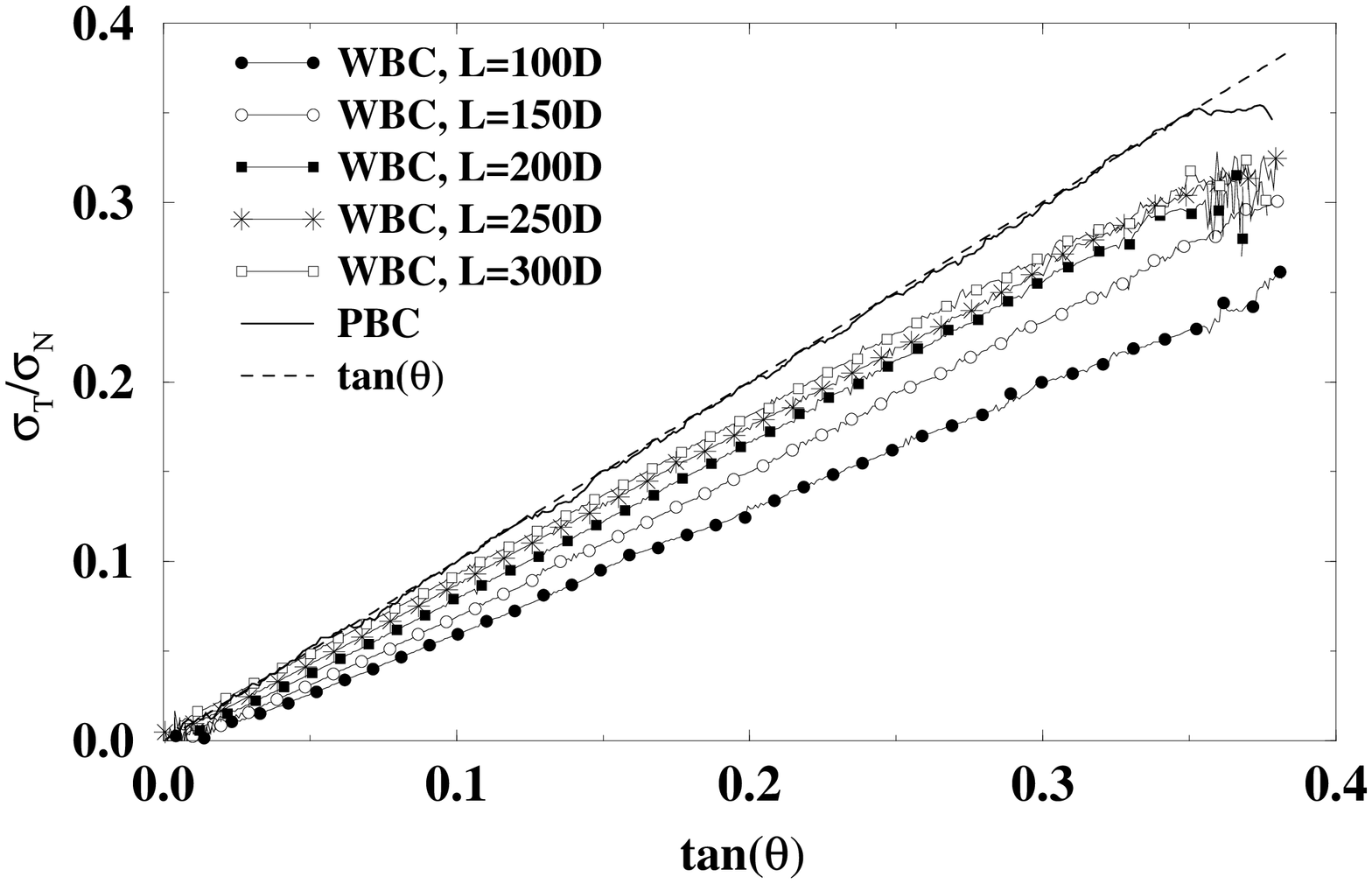,angle = -90,width =0.95\linewidth}}
\caption{Evolution of the normalized shear stress $\sigma_T/\sigma_N$ as a function of $\tan\theta$ for all WBC and PBC systems. The dotted line corresponds to the analytical fit $\sigma_T/\sigma_N = \tan(\theta)$.}
\label{Fig4}
%\end{figure}

%     Fig5.eps
%\begin{figure}
\centerline{\epsfig{file = 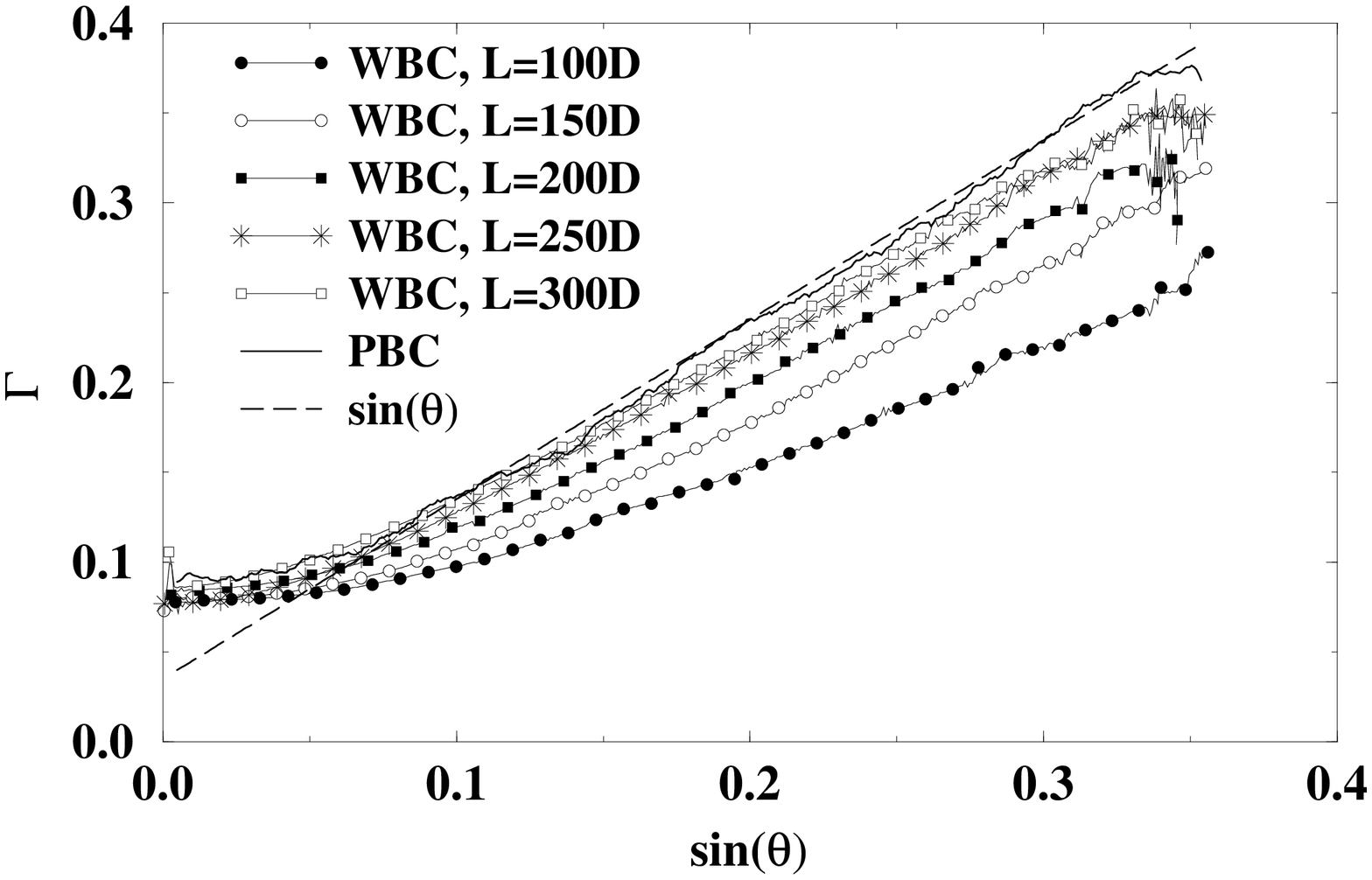,angle = -90,width =0.95\linewidth}}
\caption{Evolution of the stress ratio $\Gamma$ as a function of $\sin \theta$ for all WBC and PBC systems. The dotted line corresponds to the analytical fit $\Gamma = \sin(\theta)$.}
\label{Fig5}
\end{figure}

Figure~\ref{Fig4} displays the evolution of the normalized shear stress $\sigma_T/\sigma_N$ as a function of $\tan\theta$ for all samples, as well as the analytical fit $\sigma_T/\sigma_N = \tan \theta$ for the PBC system (see below). 
The shear stress $\sigma_T / \sigma_N$ along the bed is initially zero for all samples. It increases linearly with $\tan \theta$ but with different slopes for all samples. For all WBC systems, the slope is below that of the PBC system, but the slope increases with the length $L$ of the granular bed.
The angle of repose $\theta_c$, where surface failure is initiated, varies between $\approx 19^\circ$ and $\approx 21^\circ$ for the WBC and the PBC systems.

The same trends are observed in Figure~\ref{Fig5} which shows the stress ratio $\Gamma$ as a function of $\sin \theta$ for all samples. The stress ratio $\Gamma$ is $\simeq 0.08$ at $\theta=0$. This corresponds to a slightly anisotropic stress state of the packings in the initial state.

%     Fig4.eps

In order to evaluate the influence of the walls, let us consider Euler's equations for a medium in static equilibrium~\cite{nedderman92}.
In a reference frame attached to the box with its origin located at the left uppermost point of the bed and with its axes x and y oriented along and perpendicular to the bed (Figure~\ref{Fig1}), respectively, we have
\begin{eqnarray}
\partial_x \sigma_{xx} + \partial_y \sigma_{yx} &=& w  \sin \theta,
\label{Euler1} \\
\partial_x \sigma_{xy} + \partial_y \sigma_{yy} &=& w  \cos \theta,
\label{Euler2}
\end{eqnarray}
where $w$ is the specific weight of the medium. The stresses are zero at the free surface. Since the stresses are evaluated for the whole bed, we have $\sigma_N= \langle \sigma_{yy} \rangle$ and $\sigma_T= \langle \sigma_{yx} \rangle$, where $\langle \dots \rangle$ denotes averaging over the whole bed.

In the PBC case, the bed is translationally invariant along the free surface, so that $\partial_x \sigma_{xx} = 0$ and $\partial_x \sigma_{xy}=0$. Then, Euler's equations yield $\sigma_{yx} = w \sin \theta \ y$ and $\sigma_{yy} = w \cos \theta \ y$, so that
\begin{equation}
\frac{\sigma_{T}}{\sigma_{N}} = \frac{\sigma_{yx}}{\sigma_{yy}} = \tan \theta.
\label{eqn7}
\end{equation}
This classical result fits excellently our data in the PBC case (Figure~\ref{Fig4}).
In particular, at failure ${\sigma_{T}} / {\sigma_{N}} = \tan \theta_c$. From Eq.~\ref{eqn1} it is easy to deduce also $\Gamma$
\begin{equation}
\Gamma= \sin \theta.
\label{eqn8}
\end{equation}
This form fits well the data in the PBC case as shown in Figure~\ref{Fig5}.

\begin{figure}
\centerline{\epsfig{file = 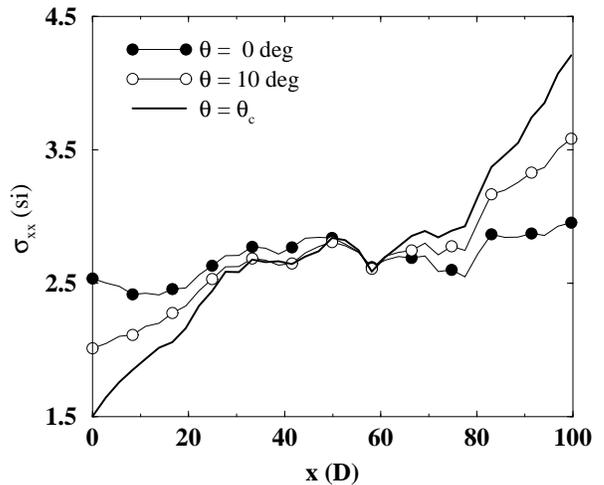,angle = -90,width =0.9\linewidth}}
\caption{Evolution of the $\sigma_{xx}$ component of the stress tensor computed over successive sections of a WBC granular bed ($L = 100D$) as a function of the distance $x$ of the section to the left wall, and for $3$ values of the slope angle $\theta$.}
\label{Fig6}
\end{figure}

In the WBC case, the translational invariance is broken in the presence of the walls. In order to close Euler's equations, we need to make an assumption about the stress state. Such an assumption represents either the outcome of a stress-stress relation, which is a standard method to close balance equations, or the history of the packing, {\it i.e.} the plastification of the bed as it is rotated from its initial state. Our data show that the main effect of the walls is to introduce a globally nonzero gradient of the $xx$ component along the bed (as a function of $x$). Figure~\ref{Fig6} shows that this effect is progressively localized in the vicinity of the walls as $\theta$ is increased. At the same time, the other components may be considered as nearly independent of $x$ up to wall effects.

In order to capture the influence of a stress gradient along the bed in an analytical approach, let us simply assume that $\sigma_{xx}$ has a constant gradient along the bed and $\partial_x \sigma_{xy} = 0$. The latter together with equation~\ref{Euler2} implies that $\sigma_{yy} = w \cos \theta \ y$. The influence of higher order gradients may be evaluated as well. But, we are concerned here only with the simplest level of description. 
Now, let us use two coefficients $k_0$ and $k_L$ to specify the boundary values of $\sigma_{xx}$:
\begin{eqnarray}
\sigma_{xx} (x=0) &=& k_0 w  \sin \theta, \label{k0} \\
\sigma_{xx} (x=L) &=& k_L w  \sin \theta. \label{kL}
\end{eqnarray}
These two coefficients might depend on $\theta$. In the PBC case, we have $k_0 = k_L =1$, and their values in the presence of the walls should reflect thus small perturbation to translational invariance.

We naturally expect that the normal stress $\sigma_{xx} (x=L)$ on the lower wall is larger than the normal stress $\sigma_{xx}(x=0)$ on the upper wall when $\theta > 0$, so that $k_L > k_0$. 
Given these boundary conditions, and since $\sigma_{xx}$ is assumed to have a constant gradient, we have
\begin{equation}
\sigma_{xx} = \left\{ k_0 + \frac{x}{L} (k_L - k_0) \right\} w \sin \theta \ y.
\label{eqn9}
\end{equation}
Then, from equation~\ref{Euler1} we get
\begin{equation}
\frac{\sigma_{T}}{\sigma_{N}} =
\tan \theta \left\{ 1 - \frac{k_L - k_0}{3} \ \frac{H}{L} \right\}.
\label{eqn10}
\end{equation}

%     Fig6.eps

%     Fig7.eps
\begin{figure}
\centerline{\epsfig{file =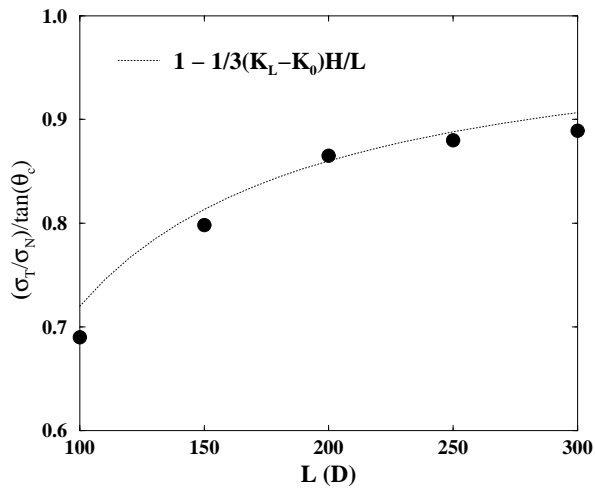,angle = -90,width =0.9\linewidth}}
\caption{Ratio $(\sigma_{T}/\sigma_{N})/\tan\theta$ for $\theta=\theta_c$ as a function of $L$ for the WBC system (black circles). The dotted line represents the analytical form given by Eq.~\protect{\ref{eqn10}}, with $k_L - k_0 \simeq 2.5$.}
\label{Fig7}
\end{figure}

This equation, with basically one fitting parameter $k_L - k_0$, captures all the features observed in Figure~\ref{Fig4}.
In particular, since $k_L - k_0  > 0$, the ratio increases with $L$ and it tends to $\tan \theta$ as $H/L \rightarrow \infty$. Figure~\ref{Fig7} shows the variation of $(\sigma_{T}/\sigma_{N})/\tan\theta$ for $\theta=\theta_c$ as a function of $L$ from the numerical data, together with the analytical form Eq.~\ref{eqn10}, which fits correctly the data with $k_L - k_0 \simeq 2.5$.

In the forthcoming sections, we focus on the PBC system, though all the observed features are basically the same in WBC systems.

%----------------------------------------------------------------
\section{Local stresses}
\label{microstress}

\subsection{Grain stress distributions}
\label{grain_stress}

\begin{figure}
\centerline{\epsfig{file = 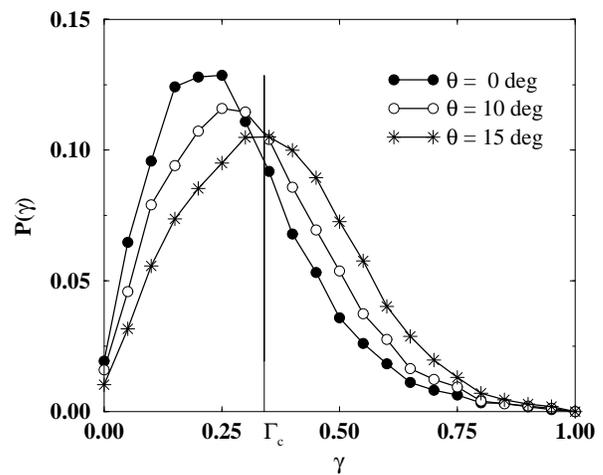,angle = -90,width = 0.9\linewidth}}
\caption{Probability density function (pdf) $P(\gamma)$ of the grainstress ratios $\gamma = q/p$ for $\theta = 0$, $\theta = 10^\circ$ and $\theta = 15^\circ$ in the PBC system. The vertical line shows the value of the critical stress ratio $\Gamma_c$ of the packing.}
\label{Fig8}
\end{figure}

\begin{figure}
\centerline{\epsfig{file = 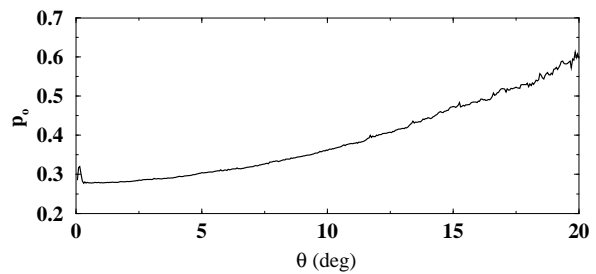,angle = -90,width =0.9\linewidth}}
\caption{Evolution of the fraction $p_o$ of overloaded grains in the PBC system as a function of the slope angle $\theta$.}
\label{Fig9}
\end{figure}

According to the definition of grain stresses (equation~\ref{eqn4}), one can attribute to each grain in the packing a stress deviator $q$ and a mean stress $p$. Figure~\ref{Fig8} displays the probability density function (pdf) $P(\gamma)$ of the grain stress ratios $\gamma = q/p$ for $\theta = 0$, $\theta = 10^\circ$ and $\theta = 15^\circ$ in the PBC system.
These distributions are quite wide and cover all possible values of $\gamma$ in the range $[0,1]$. In particular, even at low slope angles, a large fraction of the grains has a stress ratio $\gamma$ above the critical stress ratio $\Gamma_c$ of the packing (represented by a vertical line in Figure~\ref{Fig8}).
The initial fraction (at $\theta=0$) of these ``overloaded'' grains is about
$30\%$. Indeed, the grain failure thresholds (the largest permissible
 grain stress ratios) are controlled by the immediate environment of
 each grain (orientations of contact neighbours) and hence they can
 overpass the threshold $\Gamma_c$ for the packing as a whole.
This fraction grows with $\theta$ and eventually reaches $\simeq 60\%$
at $\theta_c$, as shown in Figure~\ref{Fig9}.
This suggests that the proportion of overloaded grains is a good indicator of
the ``distance'' to slope failure (an order parameter, as discussed below).

\begin{figure}
\centerline{\epsfig{file = 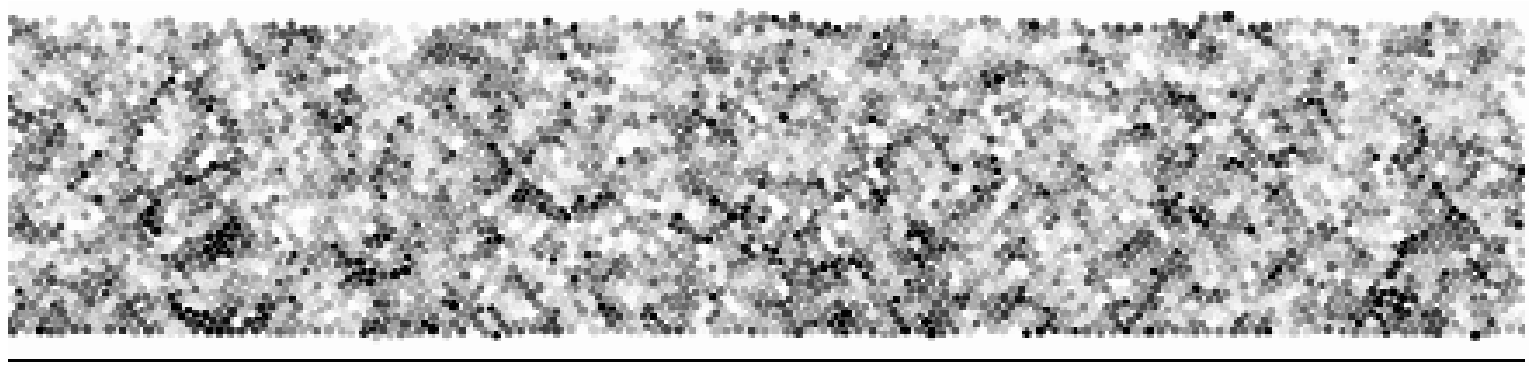 , width = \linewidth}}
\centerline{\epsfig{file = 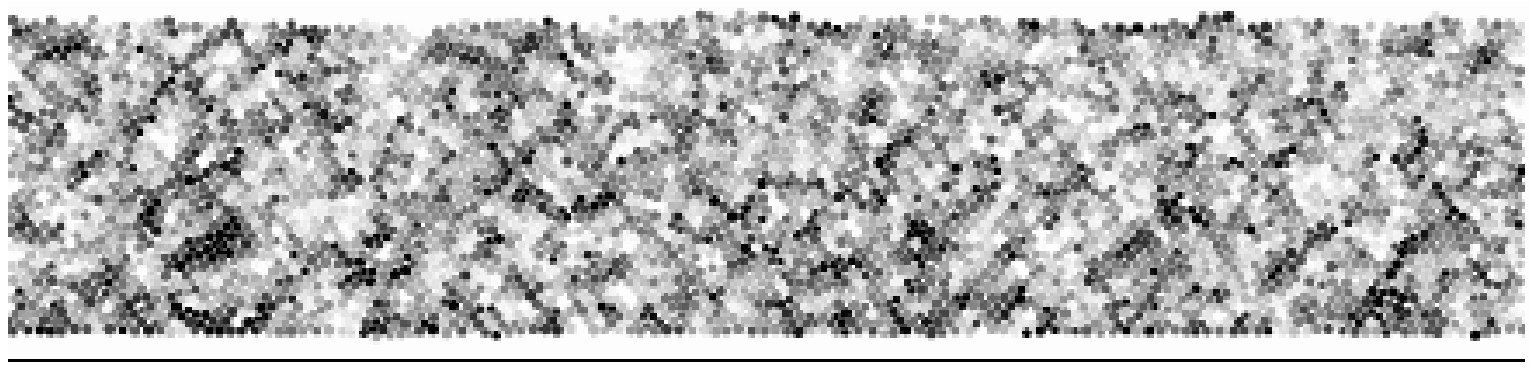 , width = \linewidth}}
\centerline{\epsfig{file = 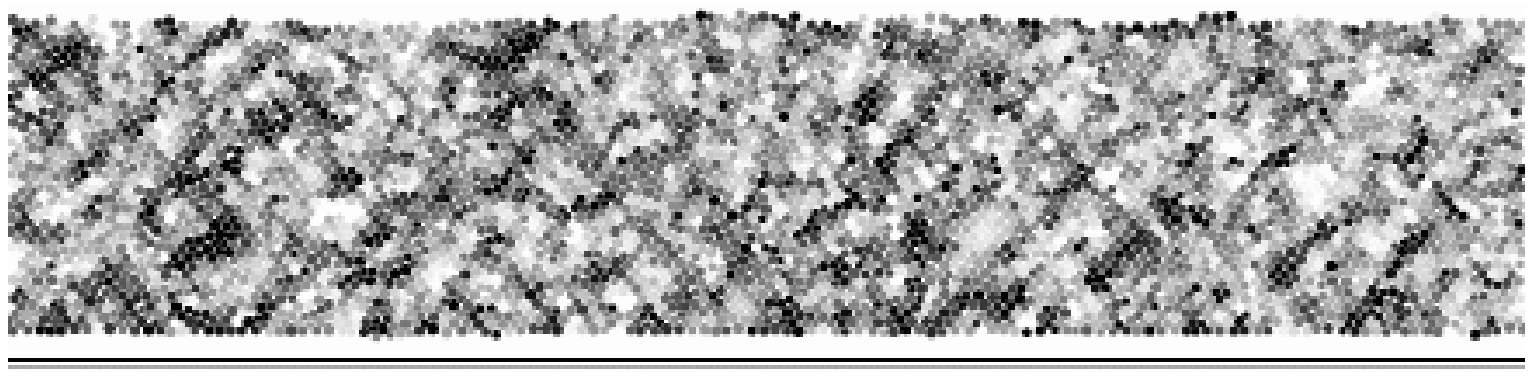 , width = \linewidth}}
\centerline{\epsfig{file = 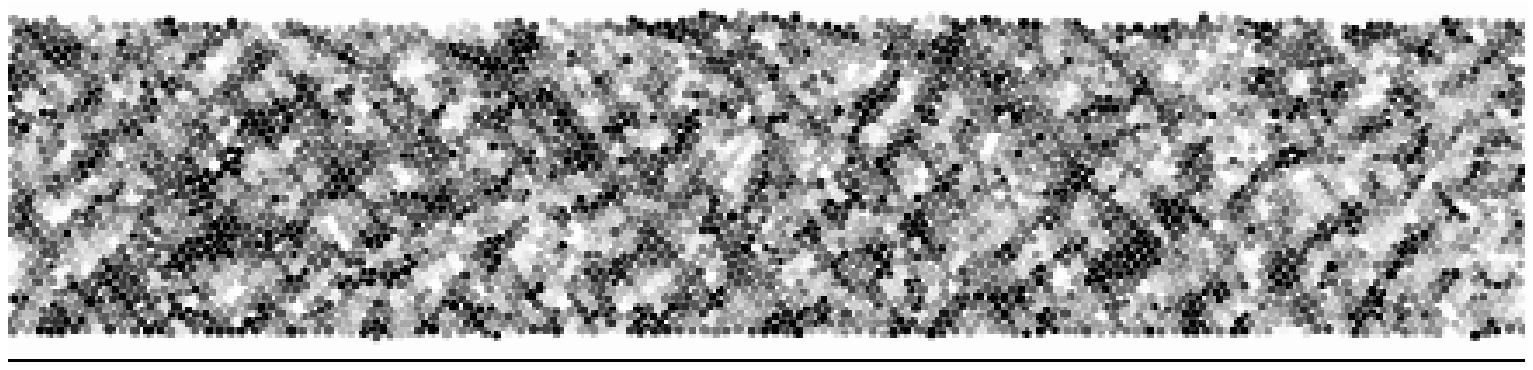 , width = \linewidth}}
\centerline{\epsfig{file = 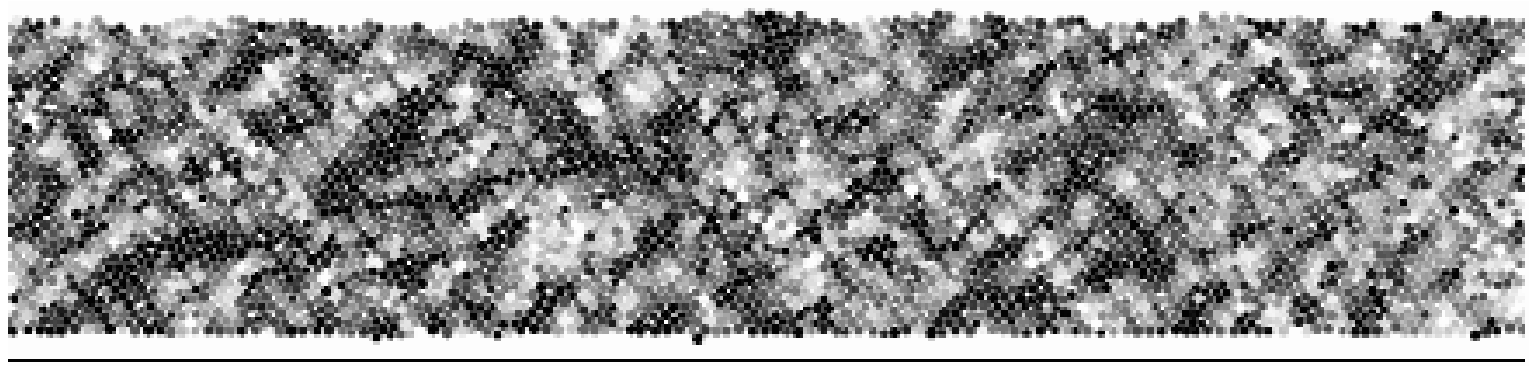 , width = \linewidth}}
\caption{Maps of the grains stress ratios $\gamma$ for successive values of $\theta$. From top to bottom, $\theta =0$, $\theta =5^\circ$, $\theta =10^\circ$, $\theta =15^\circ$ and $\theta = \theta_c$. In black are represented the grains such that $\gamma > 2\Gamma_c$, while the white grains corresponds to $\gamma < 0.05 $; intermediate values are encoded by a linear gray scale.}
\label{Map1}
\end{figure}

Figure~\ref{Map1} displays the maps of grain stress ratios $\gamma$
at different stages of the evolution using a gray level scale.
 Two cutoff values are (arbitrarily) introduced in order to improve
the contrast: The grains  with $\gamma < 0.05$ appear in white,
whereas the  grains with  $\gamma > 2 \Gamma_c$ appear in black.
Intermediate values of $\gamma$ are encoded by a linear gray level scale.
We can see on successive maps an increasing number of highly loaded
grains that remain so in spite of recurrent rearrangment phenomena
in the packing. Furthermore we observe a remarkable organization of
overloaded grains in the packing with a trend to form clusters. The map
reveals also privileged directions (at $60^\circ$ and $120^\circ$ with the
horizontal direction) corresponding with higher concentrations of
overloaded grains.

%-----FIGURES: 8,9, 10
 
%     Fig8.eps

%     Fig9.eps

%     Fig10

\subsection{A percolation process}
\label{perco}

We characterize here the clustering of overloaded grains by
analyzing their neighbourhoods. The critical stress ratio
$\Gamma_c$ is a large-scale property whereas the grain stress ratios $\gamma$
describe the grain scale stress state of the packing.
Since we have $\gamma \geq \Gamma_c$ for overloaded grains, we expect that
the average stress ratio over a larger and larger
neighbourhood of an overloaded grain will decline and eventually go
 below $\Gamma_c$. The extent of the neighbourhood for which
 $\gamma = \Gamma_c$ is thus an intermediate length scale
 (between the grain size and the system size) that
reflects the spatial correlations of overloaded grains.

\begin{figure}
\centerline{\epsfig{file = 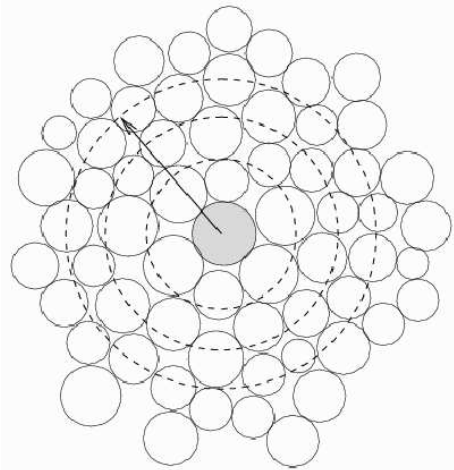 ,clip = , width = 0.7\linewidth}}
\caption{Successive circular neighbourhoods of diameter $\ell$ centered on one overloaded grain (in gray). The stress ratios $\gamma_\ell$ of the grain are computed over these neighbourhoods as a function of $\ell$.}
\label{Fig11}
\end{figure}

In order to evaluate these correlations from the data, we consider
circular neighbourhoods of diameter $\ell$ centered on
overloaded grains (Figure~\ref{Fig11}). For each grain, we
calculate  the stress ratio $\gamma_\ell$
over its circular neighbourhood as a function of $\ell$. Thereby
we determine, for each grain and at different slope angles $\theta$,
the length $\ell_c$ for which $\gamma_\ell=\Gamma_c$.
Here, we study the evolution of the maximum value $\ell_c^{max}$
and the mean value $\ell_{c}^{mean}= ({1}/{N_o})\sum_{N_o} \ell_c$
as a function of $\theta$.

\begin{figure}
\centerline{\epsfig{file = 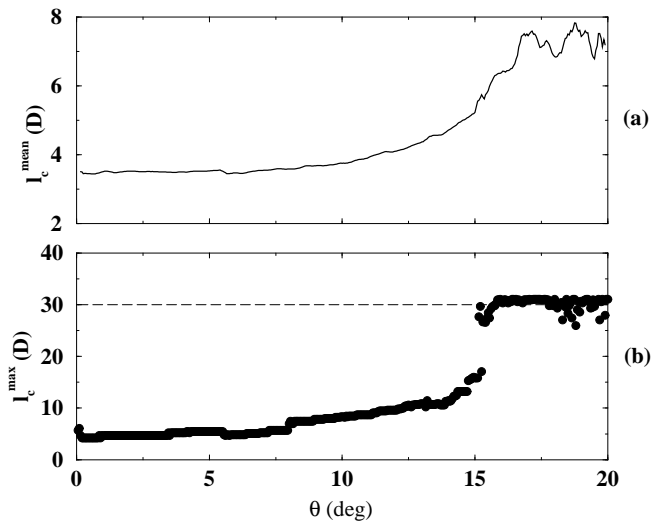,angle = -90,width = 0.99\linewidth}}
\caption{Evolution of $\ell_{c}^{mean}$ (a) and $\ell_{c}^{max}$ (b) as a function of $\theta$ (see text for definitions).}
\label{Fig12}
\end{figure}

The plots of $\ell_{c}^{mean}$ and $\ell_{c}^{max}$ as a function of $\theta$
 are displayed in Figure~\ref{Fig12}. The mean radius $\ell_{c}^{mean}$
increases very slowly from $4D$ to $5D$. Then, from $\theta=15^\circ$ onwards,
it increases rapidly to $\simeq 7D$.
At the same angle $\theta=15^\circ \equiv \theta_d$, we observe a very
spectacular  transition in the evolution of $\ell_{c}^{max}$.
The latter first increases continuously from $5D$ at $\theta=0$ to $15D$
at $\theta=\theta_d$. Here, a sudden jump brings
up $\ell_{c}^{max}$ from $15D$ to $30D$. The cutoff at $30D$ is imposed by
data processing for the evaluation of $\ell_c$ for individual grains.
However, this value is close enough to the depth $H=40D$ of
the granular bed to be interpreted as corresponding to a system
 size correlation.

It is interesting to map the growth of $\ell_{c}^{max}$ and
its nearly discontinuous change at $\theta = \theta_d$
into a percolation process of the overloaded grains.
Indeed, $\ell_{c}^{max}$ increases with $\theta$ because
of the growing number of overloaded grains (Figure~\ref{Fig9}).
Each overloaded grain represents an active site which can belong to
a nearest-neighbour cluster. In this picture, $\ell_{c}^{max}$
corresponds to the size of the largest cluster of active sites,
while $\ell_{c}^{mean}$ represents the mean cluster size.
In this sense, the transition at $\theta = \theta_d$ is a percolation
transition where the largest cluster size diverges (spanning the whole system).
The fraction of active sites at the transition is $p_o=0.45$
(Figure~\ref{Fig9}). Interestingly, this
proportion is the percolation threshold for randomly built-up structures
of non-overlapping particles in two dimensions~\cite{frisch61,sykes64,scher70}.

\begin{figure}
\centerline{\epsfig{file = 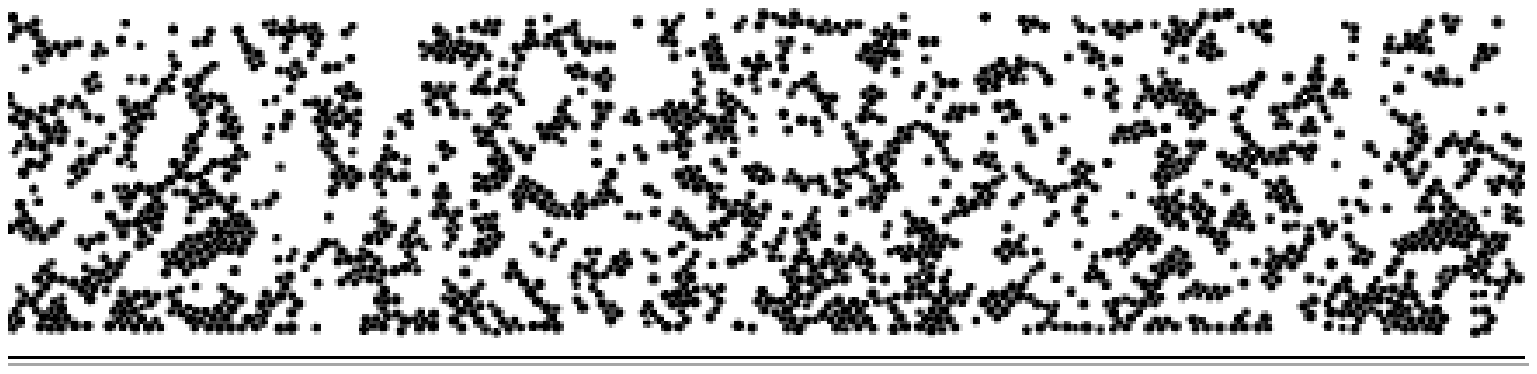,clip =, width = \linewidth}}
\centerline{\epsfig{file = 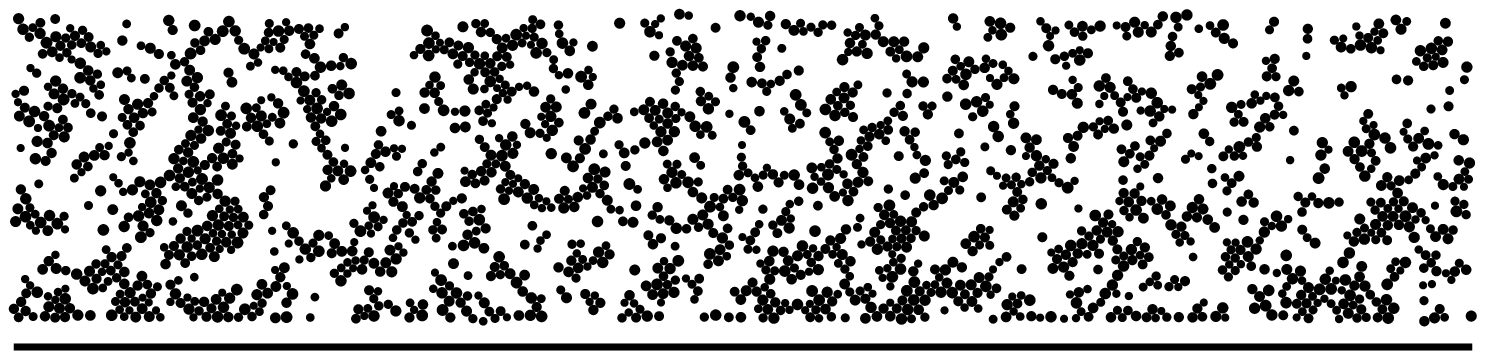,clip =, width = \linewidth}}
\centerline{\epsfig{file = 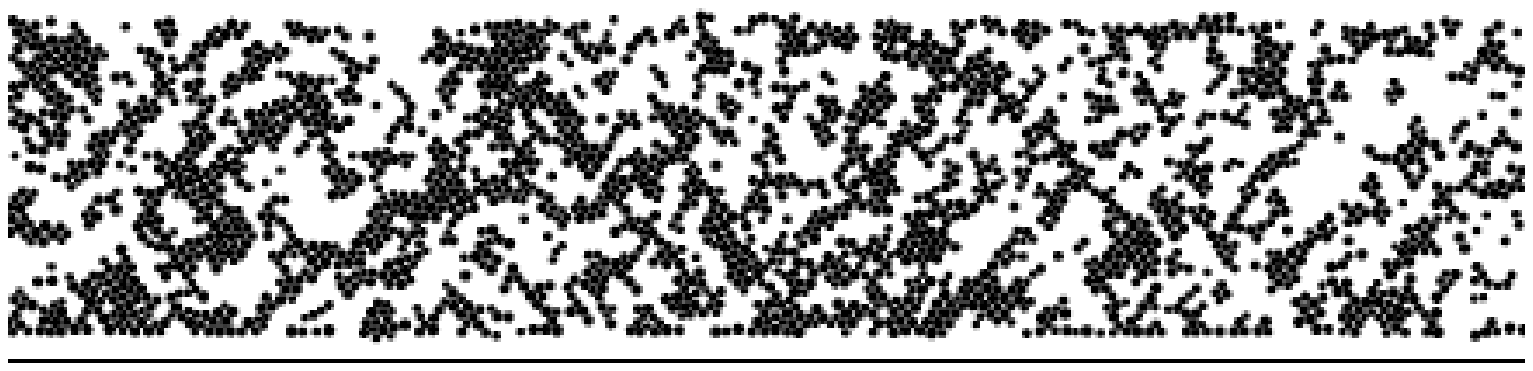,clip =, width = \linewidth}}
\centerline{\epsfig{file = 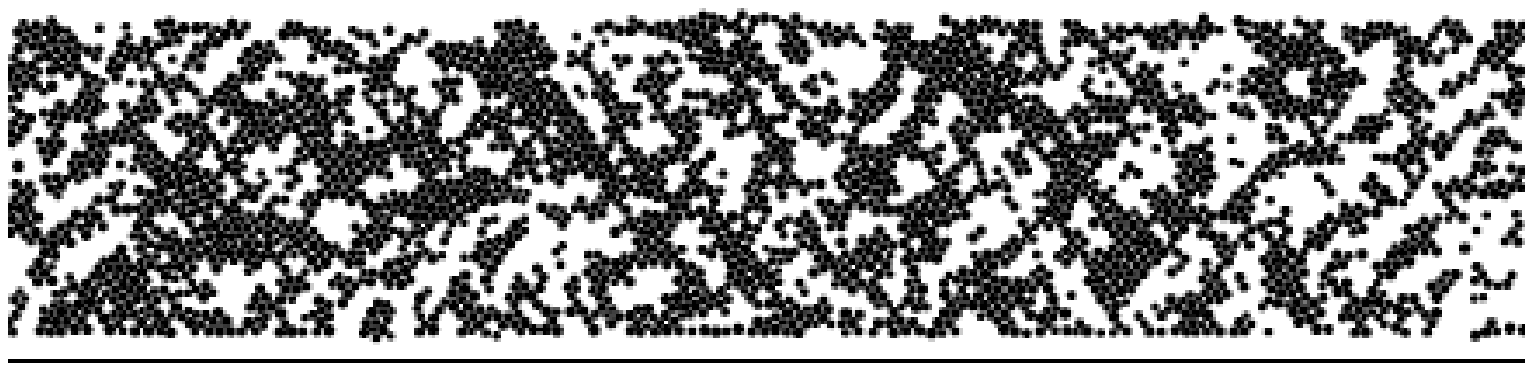, clip =,width = \linewidth}}
\centerline{\epsfig{file = 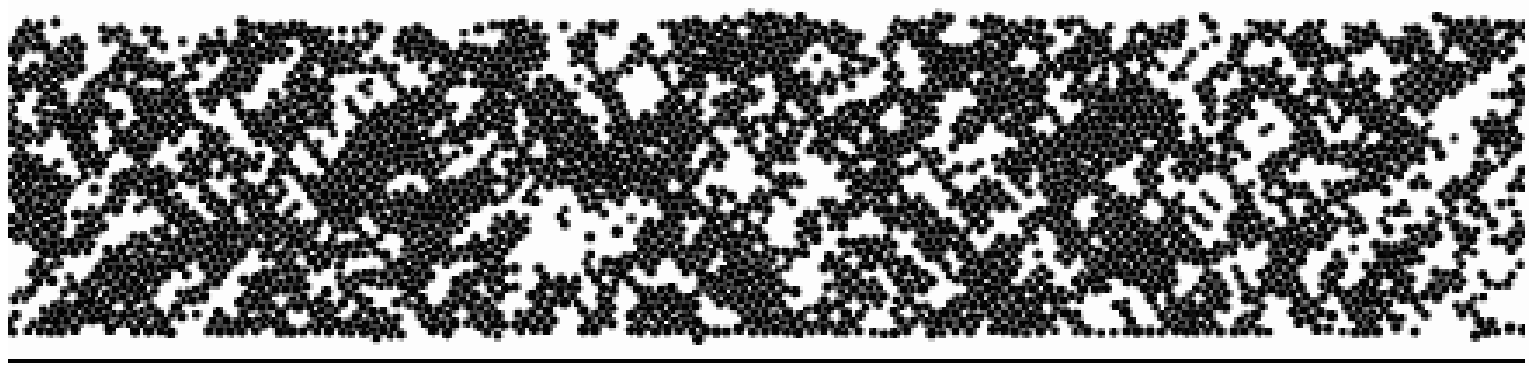, clip =,width = \linewidth}}
\caption{Maps of the overloaded grains (in black) for successive values of $\theta$. From top to bottom, $\theta =0$, $\theta =5^\circ$, $\theta =10^\circ$, $\theta =15^\circ$ and $\theta = \theta_c$. They form growing clusters which eventually span the packing from top to bottom.}
\label{Map2}
\end{figure}

Figure~\ref{Map2} shows successive snapshots of the packing in the
frame attached to the simulation box for an increasing slope angle
with overloaded grains filled in black. We clearly
observe the clusters of overloaded sites that grow in size and coalesce
already at $\theta=15^\circ$ to form a connex network spanning the
packing from top to bottom. Beyond this point, the overloaded grains
 continue to percolate (while their number increases) until
slope failure occurs.

Note that the circular shape of our control volume
does not differentiate space directions. However, as mentionned
before, because of our tight grain size distributions,
there are privileged directions at which a larger concentration of
overloaded grains can be observed (Figure~\ref{Map1}). This means that,
if the percolation of overloaded grains was analyzed for different
space directions, we would get a ``directed'' percolation at
those privileged directions for a slope angle slightly below
$\theta=15^\circ$.

%-----FIGURES: 11,12,13

%     Fig11.eps

%     Fig12.eps

%     Fig13

\section{Transition to coherent shearing}
\label{transition}

\begin{figure}
\centerline{\epsfig{file = 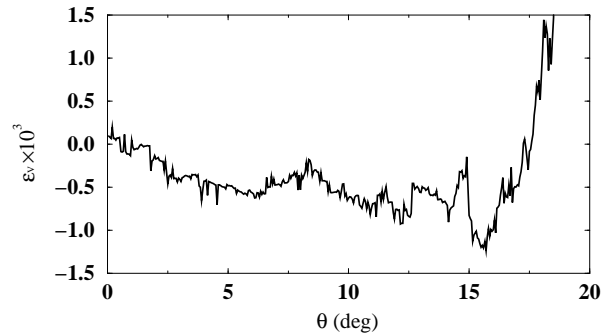,angle = -90,width =0.9\linewidth}}
\caption{Evolution of the volumetric strain $\varepsilon_V = (V-V_0)/V_0$, $V_0$ being the initial volume, as a function of $\theta$ in the PBC system. }
\label{Fig14}
\end{figure}

The evolution of the packing stress ratio $\Gamma$
with $\theta$ (Figure~\ref{Fig5}) carries no apparent signature
 of a discontinuous transition at $\theta_d$. Irrespective
of boundary conditions and for different values of the aspect ratio
 $H/L$, the packing stress follows closely the loading direction up to slope
failure. Such a signature should thus appear in the volume-change behaviour
of the packing. Indeed, numerous small rearrangements in the bed
lead to small variations of the total volume $V$ of the bed.
Figure~\ref{Fig14} shows the volumetric strain
$\varepsilon_V = (V-V_0)/V_0$, where $V_0$  is the initial volume,
as a function of $\theta$ in the PBC system.
We see that the volume first decreases (negative values of $\varepsilon_V$)
with $\theta$. Then, precisely at $\theta = \theta_d$, this contractant
 behaviour transforms into a dilatant behaviour, with the volume increasing
 (positive values of $\varepsilon_V$) up to surface failure.

Since packing dilatation may occur only as a consequence of shearing,
the transition to dilatation at $\theta_d$ can be interpreted as the onset of
a stable shear mode in the packing. Before transition, the
particle rearrangements occur in a diffuse and incoherent manner in the whole
packing. This is due to the fact that, in the presence of geometrical
disorder and gravity, each grain tends to
occupy a more stable position in the packing. Such rearrangments take place
mostly in a collective way but in small volumes compared to the size of the
system~\cite{staron02}, and they do not disturb the overall stability of
 the slope (neither at the free surface nor in the bulk)
as long as they have low amplitudes (involving low rotation rates).

This behaviour is reminiscent of the
contractant behaviour observed in the first stages of a shear test
performed on soil samples~\cite{hicher}. The extent of volume reduction depends
on the
initial solid fraction of the packing.
A transition to dilatancy happens if the
initial solid fraction is above the ``critical state'' solid fraction,
{\it i.e.} the solid fraction corresponding to a state reached after
 long enough monotonous shearing~\cite{wood_book}.
This is an interesting analogy although in a granular slope the gravity
 behaves as a bulk force in contrast to shear testing conditions
where the largest stresses are exerted at the wall boundaries on the sample.
Moreover, the rotation of principal stress directions with respect to
the packing  (as in our rotating bed) is not a ``standard''
testing method for the behaviour of soil samples.

%------------Figures: 14

%      Fig14.eps

\section{Discussion and Conclusion}
\label{conclusion}

\subsection{Summary}

In this paper, we analyzed the stress states in
a granular slope composed of rigid
disks and simulated by the Contact Dynamics method.
We showed that a global stress analysis in terms of Euler's
equations for periodic boundary conditions along the free
surface fits excellently the simulation data. The influence of
side walls was also evaluated and compared to the data as
a function of the total length $L$ of the free surface.

The local stresses were described in terms of internal
moments which allow for an additive coarse-graining of
stresses up to the system size. We observed a wide
distribution of grain stress ratios involving a significant fraction of
the so-called ``overloaded'' grains,
for which the stress ratio $\gamma$ is above the ``critical''
stress ratio $\Gamma_c$, corresponding to
the Coulomb yield threshold for the whole packing.

As long as the slope angle $\theta$ is below the angle of repose $\theta_c$,
the stress ratio in a control volume centered on an
overloaded grain decreases with the size $\ell$ of the control volume.
This property allowed us to define a ``correlation'' length $\ell_c$
as the size of a control volume centered on each overloaded grains, and
for which the stress ratio is just equal to the
critical stress ratio $\Gamma_c$.
The largest length $\ell_c^{max}$ in the packing is
thus a coarse-graining length beyond which all stress ratios are below
the Coulomb yield threshold.

We found that $\ell_c^{max}$ increases as the bed is slowly tilted towards
its angle of repose; at an angle $\theta_d$ below the
angle of repose $\ell_c^{max}$ undergoes a sudden jump to
a length comparable to the depth of the bed.
We argued that this evolution of local stresses can
be mapped into a percolation process of the overloaded grains.
In this picture, the discontinuous increase of the coarse-graining length
at $\theta_d$ corresponds to a percolation transition. Remarkably, this
transition coincides with the onset of dilatation in the packing.

\subsection{Discussion}

The above results reveal an unexpectedly rich scaling of stresses in a
granular bed in the vicinity of slope failure. In particular,
they are consistent with the advent of a ``phase transition''
at $\theta_d$. Upon transition,
the stress inhomogeneities, in the form of clustering of overloaded
grains, extend to the whole packing. At the same time,
the grain rearrangements turn from an incoherent behaviour
into coherent dilatant shearing.
This transition is all the more interesting that
it occurs far enough from the angle of repose $\theta_c$ to
be considered as a strong precursor of the incoming surface instability.

Although the transition angle $\theta_d$ appears here in the course of a
``quasi-static'' evolution of the slope, it is still tempting to identify
$\theta_d$ with the {\em dynamic} angle of repose, {\it i.e.}
the angle reached when the slope comes to rest after failure.
This is an appealing interpretation in that it relates the
dynamic angle of repose to a {\em static} property of the packing.
In other words, the order parameter in this description is the fraction of
 overloaded grains.
Further simulations are necessary to check this conjecture, but let us
 simply consider here an argument in this direction.

In the presence of long-range correlations or large clusters of overloaded
grains for the angles $\theta$ in the interval
$[\theta_d,\theta_c]$, the granular bed should be
metastable (sensitive to perturbations).
This is consistent with experimental measurements
of the size of the perturbation (number of grains added at a
 point on top of the bed) required to trigger an avalanche,
 which is shown to increase as a function of the distance
$\delta \theta =\theta_c -\theta$ to the angle of repose~\cite{daerr00}.
A surface flow triggered at $\theta_c$ acts as a perturbation that
will continue to exist as long as the slope is metastable.
Hence, it will not stop until $\theta$ declines below $\theta_d$.
This means that the avalanche destroys long-range
correlations and takes the slope back to a ``disorganized'' state.

It is worth noting that we did not observe such a clear transition
for the pair correlation functions of contact forces and grain stresses
at $\theta_d$. In these cases, the correlation length remains of the order of
a few grain diameters all along the simulation. However, as shown in detail
in~\cite{staron02}, the evolution of ``critical contacts'', {\it i.e.}
contacts where the friction is fully mobilized, is consistent with
a percolation-like process (of critical contacts),
leading to a power-law increase of the corresponding correlation length
from $\theta_d$ onward and a divergence
of the correlation length at $\theta_c$. This behaviour is
plausible in view of the onset of coherent shearing at $\theta_d$, as shown
in this paper.

This suggests that the evolution of a granular bed
is best described in terms of overloaded grains in the range
$[0,\theta_d]$ and in terms of critical contacts in the
range $[\theta_d,\theta_c]$.
Indeed, at slope angles below $\theta_d$,
the fraction of overloaded grains is a good parameter in that
it increases strongly with the slope angle and it  controls the percolation
transition at $\theta_d$. In this interval, the friction mobilization
is rather chaotic and incoherent, and the fraction of critical contacts
increases
 very slowly with the slope angle~\cite{staron02}. At slope angles above
 $\theta_d$, due to coherent shearing, the friction forces are
increasingly and coherently
mobilized so that the evolution can be described in terms of
critical contacts (the order parameter being the fraction of these contacts).
This evolution leads to a
power-law divergence of the correlation length for critical contacts at
$\theta_c$.

The dilatation of the bed provides another interesting insight
into the behaviour of a granular bed in the metastable range.
The avalanche can not occur unless the bed dilates. Since the bed begins
to dilate at $\theta_d$, the angle $\theta_c - \theta_d$ should
be interpreted as the ``dilatation angle'' of the granular material
 when plastified as a result of rearrangements induced by the rotation
 of the bed~\cite{nedderman92}.
Hence, in a macroscopic approach, the static angle of repose $\theta_c$ and
the transition angle  $\theta_d$, interpreted as the dynamic angle of repose,
may be described in terms of the Coulomb yield criterion and the
dilatation angle, respectively.

\acknowledgements

We acknowledge very fruitful discussions with J. J. Moreau.
This work was supported by the Marie Curie European Fellowship FP6 Program.

%\bibliography{Ref}

\begin{thebibliography}{40}
\expandafter\ifx\csname natexlab\endcsname\relax\def\natexlab#1{#1}\fi
\expandafter\ifx\csname bibnamefont\endcsname\relax
  \def\bibnamefont#1{#1}\fi
\expandafter\ifx\csname bibfnamefont\endcsname\relax
  \def\bibfnamefont#1{#1}\fi
\expandafter\ifx\csname citenamefont\endcsname\relax
  \def\citenamefont#1{#1}\fi
\expandafter\ifx\csname url\endcsname\relax
  \def\url#1{\texttt{#1}}\fi
\expandafter\ifx\csname urlprefix\endcsname\relax\def\urlprefix{URL }\fi
\providecommand{\bibinfo}[2]{#2}
\providecommand{\eprint}[2][]{\url{#2}}

\bibitem[{\citenamefont{Coulomb}(1773)}]{coulomb73}
\bibinfo{author}{\bibfnamefont{C.}~\bibnamefont{Coulomb}},
  \bibinfo{journal}{Acad. Roy. Sci. Math. Phys.} \textbf{\bibinfo{volume}{7}},
  \bibinfo{pages}{343} (\bibinfo{year}{1773}).

\bibitem[{\citenamefont{Drucker}(1954)}]{drucker54}
\bibinfo{author}{\bibfnamefont{D.}~\bibnamefont{Drucker}},
  \bibinfo{journal}{Journal of Applied Mechanics} pp. \bibinfo{pages}{71--74}
  (\bibinfo{year}{1954}), \bibinfo{note}{analyse limite}.

\bibitem[{\citenamefont{Wood}(1990)}]{wood_book}
\bibinfo{author}{\bibfnamefont{D.}~\bibnamefont{Wood}},
  \emph{\bibinfo{title}{Soil behaviour and critical state soil mechanics}}
  (\bibinfo{publisher}{Cambridge University Press},
  \bibinfo{address}{Cambridge, England}, \bibinfo{year}{1990}).

\bibitem[{\citenamefont{Oda and Iwashita}(1999)}]{oda99}
\bibinfo{editor}{\bibfnamefont{M.}~\bibnamefont{Oda}} \bibnamefont{and}
  \bibinfo{editor}{\bibfnamefont{K.}~\bibnamefont{Iwashita}}, eds.,
  \emph{\bibinfo{title}{Mechanics of Granular Materials}}
  (\bibinfo{publisher}{A. A. Balkema}, \bibinfo{address}{Rotterdam},
  \bibinfo{year}{1999}).

\bibitem[{\citenamefont{Jaeger and Nagel}(1996)}]{jaeger_nagel96}
\bibinfo{author}{\bibfnamefont{H.}~\bibnamefont{Jaeger}} \bibnamefont{and}
  \bibinfo{author}{\bibfnamefont{S.}~\bibnamefont{Nagel}},
  \bibinfo{journal}{Reviews of Modern Physics} \textbf{\bibinfo{volume}{68}},
  \bibinfo{pages}{1259} (\bibinfo{year}{1996}).

\bibitem[{\citenamefont{de~Gennes}(1999)}]{degennes99}
\bibinfo{author}{\bibfnamefont{P.}~\bibnamefont{de~Gennes}},
  \bibinfo{journal}{Reviews od Modern Physics}
  \textbf{\bibinfo{volume}{71}}(\bibinfo{number}{2}), \bibinfo{pages}{S374}
  (\bibinfo{year}{1999}).

\bibitem[{\citenamefont{Jaeger et~al.}(1989)\citenamefont{Jaeger, Liu, and
  Nagel}}]{jaeger89}
\bibinfo{author}{\bibfnamefont{H.~M.} \bibnamefont{Jaeger}},
  \bibinfo{author}{\bibfnamefont{C.}~\bibnamefont{Liu}}, \bibnamefont{and}
  \bibinfo{author}{\bibfnamefont{S.~R.} \bibnamefont{Nagel}},
  \bibinfo{journal}{Phys. Rev. Lett.}
  \textbf{\bibinfo{volume}{62}}(\bibinfo{number}{1}), \bibinfo{pages}{40}
  (\bibinfo{year}{1989}).

\bibitem[{\citenamefont{Darve and Laouafa}(2000)}]{darve00}
\bibinfo{author}{\bibfnamefont{F.}~\bibnamefont{Darve}} \bibnamefont{and}
  \bibinfo{author}{\bibfnamefont{F.}~\bibnamefont{Laouafa}},
  \bibinfo{journal}{Mechanics of cohesive-frictional materials}
  \textbf{\bibinfo{volume}{5}}, \bibinfo{pages}{627} (\bibinfo{year}{2000}).

\bibitem[{\citenamefont{Daerr}(2000)}]{daerr00}
\bibinfo{author}{\bibfnamefont{A.}~\bibnamefont{Daerr}}, Ph.D. thesis,
  \bibinfo{school}{Universit\'e Denis Diderot Paris 7} (\bibinfo{year}{2000}).

\bibitem[{\citenamefont{Quartier et~al.}(2000)\citenamefont{Quartier,
  Andreotti, Douady, and Daerr}}]{quartier00}
\bibinfo{author}{\bibfnamefont{L.}~\bibnamefont{Quartier}},
  \bibinfo{author}{\bibfnamefont{B.}~\bibnamefont{Andreotti}},
  \bibinfo{author}{\bibfnamefont{S.}~\bibnamefont{Douady}}, \bibnamefont{and}
  \bibinfo{author}{\bibfnamefont{A.}~\bibnamefont{Daerr}},
  \bibinfo{journal}{Physical Review E}
  \textbf{\bibinfo{volume}{62}}(\bibinfo{number}{6}), \bibinfo{pages}{8299}
  (\bibinfo{year}{2000}).

\bibitem[{\citenamefont{Alonso et~al.}(1998)\citenamefont{Alonso, Hovi, and
  Herrmann}}]{alonso98}
\bibinfo{author}{\bibfnamefont{J.~J.} \bibnamefont{Alonso}},
  \bibinfo{author}{\bibfnamefont{J.-P.} \bibnamefont{Hovi}}, \bibnamefont{and}
  \bibinfo{author}{\bibfnamefont{H.~J.} \bibnamefont{Herrmann}},
  \bibinfo{journal}{Phys. Rev. E} \textbf{\bibinfo{volume}{58}},
  \bibinfo{pages}{672} (\bibinfo{year}{1998}).

\bibitem[{\citenamefont{tre}(2002)}]{lemaitre02}
\bibinfo{author}{\bibfnamefont{A.~L.} \bibnamefont{tre}},
  \bibinfo{journal}{Physical Review Letters}
  \textbf{\bibinfo{volume}{89}}(\bibinfo{number}{6}), \bibinfo{pages}{064303}
  (\bibinfo{year}{2002}).

\bibitem[{\citenamefont{Staron et~al.}(2002)\citenamefont{Staron, Vilotte, and
  Radjai}}]{staron02}
\bibinfo{author}{\bibfnamefont{L.}~\bibnamefont{Staron}},
  \bibinfo{author}{\bibfnamefont{J.-P.} \bibnamefont{Vilotte}},
  \bibnamefont{and} \bibinfo{author}{\bibfnamefont{F.}~\bibnamefont{Radjai}},
  \bibinfo{journal}{Phys. Rev. Lett.} \textbf{\bibinfo{volume}{89}},
  \bibinfo{pages}{204302} (\bibinfo{year}{2002}).

\bibitem[{\citenamefont{Oda}(1972)}]{oda72a}
\bibinfo{author}{\bibfnamefont{M.}~\bibnamefont{Oda}}, \bibinfo{journal}{Soils
  and foundations} \textbf{\bibinfo{volume}{12}}, \bibinfo{pages}{17}
  (\bibinfo{year}{1972}).

\bibitem[{\citenamefont{Oda et~al.}(1980)\citenamefont{Oda, Koshini, and
  Nemat-Nasser}}]{oda80}
\bibinfo{author}{\bibfnamefont{M.}~\bibnamefont{Oda}},
  \bibinfo{author}{\bibfnamefont{J.}~\bibnamefont{Koshini}}, \bibnamefont{and}
  \bibinfo{author}{\bibfnamefont{S.}~\bibnamefont{Nemat-Nasser}},
  \bibinfo{journal}{Geotechnique} \textbf{\bibinfo{volume}{30}},
  \bibinfo{pages}{479} (\bibinfo{year}{1980}).

\bibitem[{\citenamefont{Rothenburg and Bathurst}(1989)}]{rothenburg89}
\bibinfo{author}{\bibfnamefont{L.}~\bibnamefont{Rothenburg}} \bibnamefont{and}
  \bibinfo{author}{\bibfnamefont{R.~J.} \bibnamefont{Bathurst}},
  \bibinfo{journal}{Geotechnique} \textbf{\bibinfo{volume}{39}},
  \bibinfo{pages}{601} (\bibinfo{year}{1989}).

\bibitem[{\citenamefont{Cambou}(1993)}]{cambou93}
\bibinfo{author}{\bibfnamefont{B.}~\bibnamefont{Cambou}}, in
  \emph{\bibinfo{booktitle}{Powders and Grains 93}}, edited by
  \bibinfo{editor}{\bibfnamefont{C.}~\bibnamefont{Thornton}}
  (\bibinfo{publisher}{A. A. Balkema}, \bibinfo{address}{Amsterdam},
  \bibinfo{year}{1993}), pp. \bibinfo{pages}{73--86}.

\bibitem[{\citenamefont{Radjai}(1998)}]{radjai98}
\bibinfo{author}{\bibfnamefont{F.}~\bibnamefont{Radjai}}, in
  \emph{\bibinfo{booktitle}{Physics of dry granular media - NATO ASI Series
  E350}}, edited by \bibinfo{editor}{\bibfnamefont{H.~J.}
  \bibnamefont{Herrmann}}, \bibinfo{editor}{\bibfnamefont{J.-P.}
  \bibnamefont{Hovi}}, \bibnamefont{and}
  \bibinfo{editor}{\bibfnamefont{S.}~\bibnamefont{Luding}}
  (\bibinfo{publisher}{Kluwer Academic Publishers},
  \bibinfo{address}{Dordrecht}, \bibinfo{year}{1998}), pp.
  \bibinfo{pages}{305--311}.

\bibitem[{\citenamefont{Radjai and Roux}(2001)}]{radjai01a}
\bibinfo{author}{\bibfnamefont{F.}~\bibnamefont{Radjai}} \bibnamefont{and}
  \bibinfo{author}{\bibfnamefont{S.}~\bibnamefont{Roux}}, in
  \emph{\bibinfo{booktitle}{Powders and Grains 2001}}, edited by
  \bibinfo{editor}{\bibfnamefont{Y.}~\bibnamefont{Kishino}}
  (\bibinfo{publisher}{A. A. Balkema}, \bibinfo{address}{Amsterdam},
  \bibinfo{year}{2001}), pp. \bibinfo{pages}{21--24}.

\bibitem[{\citenamefont{Howell et~al.}(1999)\citenamefont{Howell, Berhinger,
  and Veje}}]{howell99}
\bibinfo{author}{\bibfnamefont{D.}~\bibnamefont{Howell}},
  \bibinfo{author}{\bibfnamefont{R.~P.} \bibnamefont{Berhinger}},
  \bibnamefont{and} \bibinfo{author}{\bibfnamefont{C.}~\bibnamefont{Veje}},
  \bibinfo{journal}{Phys. Rev. Lett} \textbf{\bibinfo{volume}{82}},
  \bibinfo{pages}{5241} (\bibinfo{year}{1999}).

\bibitem[{\citenamefont{Darve and Laouafa}(1999)}]{darve99}
\bibinfo{author}{\bibfnamefont{F.}~\bibnamefont{Darve}} \bibnamefont{and}
  \bibinfo{author}{\bibfnamefont{F.}~\bibnamefont{Laouafa}}, in
  \emph{\bibinfo{booktitle}{Numerical Models in Geomaterials}}, edited by
  \bibinfo{editor}{\bibnamefont{Prande}},
  \bibinfo{editor}{\bibnamefont{Pietruszczak}}, \bibnamefont{and}
  \bibinfo{editor}{\bibnamefont{Schweiger}} (\bibinfo{publisher}{Balkema},
  \bibinfo{address}{Rotterdam}, \bibinfo{year}{1999}), pp.
  \bibinfo{pages}{85--90}.

\bibitem[{\citenamefont{Roux and Radjai}(2001)}]{roux01}
\bibinfo{author}{\bibfnamefont{S.}~\bibnamefont{Roux}} \bibnamefont{and}
  \bibinfo{author}{\bibfnamefont{F.}~\bibnamefont{Radjai}}, in
  \emph{\bibinfo{booktitle}{Mechanics for a New Millennium}}, edited by
  \bibinfo{editor}{\bibfnamefont{H.}~\bibnamefont{Aref}} \bibnamefont{and}
  \bibinfo{editor}{\bibfnamefont{J.}~\bibnamefont{Philips}}
  (\bibinfo{publisher}{Kluwer Acad. Pub.}, \bibinfo{address}{Netherlands},
  \bibinfo{year}{2001}), pp. \bibinfo{pages}{181--196}.

\bibitem[{\citenamefont{Moreau}(1997)}]{moreau97}
\bibinfo{author}{\bibfnamefont{J.}~\bibnamefont{Moreau}}, in
  \emph{\bibinfo{booktitle}{Friction, Arching, Contact Dynamics}}
  (\bibinfo{publisher}{World Scientific}, \bibinfo{address}{Singapore},
  \bibinfo{year}{1997}), pp. \bibinfo{pages}{233--247}.

\bibitem[{\citenamefont{Radjai and Roux}(2002)}]{radjai02}
\bibinfo{author}{\bibfnamefont{F.}~\bibnamefont{Radjai}} \bibnamefont{and}
  \bibinfo{author}{\bibfnamefont{S.}~\bibnamefont{Roux}},
  \bibinfo{journal}{Phys. Rev. Lett.}
  \textbf{\bibinfo{volume}{89}}(\bibinfo{number}{6}), \bibinfo{pages}{064302}
  (\bibinfo{year}{2002}).

\bibitem[{\citenamefont{Cundall and Stack}(1979)}]{cundall79}
\bibinfo{author}{\bibfnamefont{P.}~\bibnamefont{Cundall}} \bibnamefont{and}
  \bibinfo{author}{\bibfnamefont{O.}~\bibnamefont{Stack}},
  \bibinfo{journal}{Geotechnique}
  \textbf{\bibinfo{volume}{29}}(\bibinfo{number}{1}), \bibinfo{pages}{47}
  (\bibinfo{year}{1979}).

\bibitem[{\citenamefont{Herrmann}(1992)}]{herrmann92}
\bibinfo{author}{\bibfnamefont{H.}~\bibnamefont{Herrmann}},
  \bibinfo{journal}{Physica A} \textbf{\bibinfo{volume}{191}},
  \bibinfo{pages}{263} (\bibinfo{year}{1992}).

\bibitem[{\citenamefont{Luding}(1998)}]{luding98c}
\bibinfo{author}{\bibfnamefont{S.}~\bibnamefont{Luding}}, in
  \emph{\bibinfo{booktitle}{Physics of dry granular media - NATO ASI Series
  E350}}, edited by \bibinfo{editor}{\bibfnamefont{H.~J.}
  \bibnamefont{Herrmann}}, \bibinfo{editor}{\bibfnamefont{J.-P.}
  \bibnamefont{Hovi}}, \bibnamefont{and}
  \bibinfo{editor}{\bibfnamefont{S.}~\bibnamefont{Luding}}
  (\bibinfo{publisher}{Kluwer Academic Publishers},
  \bibinfo{address}{Dordrecht}, \bibinfo{year}{1998}), pp.
  \bibinfo{pages}{285--304}.

\bibitem[{\citenamefont{Moreau}(1988)}]{moreau88}
\bibinfo{author}{\bibfnamefont{J.}~\bibnamefont{Moreau}}, in
  \emph{\bibinfo{booktitle}{Nonsmooth mechanics and applications}}, edited by
  \bibinfo{editor}{\bibfnamefont{J.}~\bibnamefont{Moreau}} \bibnamefont{and}
  \bibinfo{editor}{\bibfnamefont{P.}~\bibnamefont{Panagiotopoulos}}
  (\bibinfo{publisher}{Springer Verlag}, \bibinfo{year}{1988}).

\bibitem[{\citenamefont{Jean and Moreau}(1992{\natexlab{a}})}]{jean92}
\bibinfo{author}{\bibfnamefont{M.}~\bibnamefont{Jean}} \bibnamefont{and}
  \bibinfo{author}{\bibfnamefont{J.}~\bibnamefont{Moreau}}, in
  \emph{\bibinfo{booktitle}{Proc. of Contact Mech. Int. Symp.}}, edited by
  \bibinfo{editor}{\bibfnamefont{A.}~\bibnamefont{Curnier}}
  (\bibinfo{year}{1992}{\natexlab{a}}), pp. \bibinfo{pages}{31--48}.

\bibitem[{\citenamefont{Jean and Moreau}(1992{\natexlab{b}})}]{jean_moreau92}
\bibinfo{author}{\bibfnamefont{M.}~\bibnamefont{Jean}} \bibnamefont{and}
  \bibinfo{author}{\bibfnamefont{J.}~\bibnamefont{Moreau}}, in
  \emph{\bibinfo{booktitle}{Proceedings of Contact Mechanics International
  Symposium}} (\bibinfo{publisher}{Presses Polytechniques et Universitaires
  Romandes}, \bibinfo{address}{Lausanne, Switzerland},
  \bibinfo{year}{1992}{\natexlab{b}}), pp. \bibinfo{pages}{31--48}.

\bibitem[{\citenamefont{Moreau}(1994)}]{moreau94}
\bibinfo{author}{\bibfnamefont{J.}~\bibnamefont{Moreau}},
  \bibinfo{journal}{European Journal of Mechanics A/Solids}
  \textbf{\bibinfo{volume}{supp.}}(\bibinfo{number}{4}), \bibinfo{pages}{93}
  (\bibinfo{year}{1994}).

\bibitem[{\citenamefont{Jean}(1995)}]{jean95}
\bibinfo{author}{\bibfnamefont{M.}~\bibnamefont{Jean}},
  \emph{\bibinfo{title}{Frictional contact in rigid or deformable bodies:
  numerical simulation of geomaterials}} (\bibinfo{publisher}{A.P.S. Salvadurai
  J.M. Boulon, Elsevier Science Publisher, Amsterdam}, \bibinfo{year}{1995}),
  pp. \bibinfo{pages}{463--486}.

\bibitem[{\citenamefont{Christoffersen
  et~al.}(1981)\citenamefont{Christoffersen, Mehrabadi, and
  Nemat-Nasser}}]{christoffersen81}
\bibinfo{author}{\bibfnamefont{J.}~\bibnamefont{Christoffersen}},
  \bibinfo{author}{\bibfnamefont{M.}~\bibnamefont{Mehrabadi}},
  \bibnamefont{and}
  \bibinfo{author}{\bibfnamefont{S.}~\bibnamefont{Nemat-Nasser}},
  \bibinfo{journal}{J. Appl. Mech.} \textbf{\bibinfo{volume}{48}},
  \bibinfo{pages}{339} (\bibinfo{year}{1981}).

\bibitem[{\citenamefont{Kruyt and Rothenburg}(1996)}]{kruyt96}
\bibinfo{author}{\bibfnamefont{N.~P.} \bibnamefont{Kruyt}} \bibnamefont{and}
  \bibinfo{author}{\bibfnamefont{L.}~\bibnamefont{Rothenburg}},
  \bibinfo{journal}{J. App. Mech.} \textbf{\bibinfo{volume}{118}},
  \bibinfo{pages}{706} (\bibinfo{year}{1996}).

\bibitem[{\citenamefont{Goddard}(1998)}]{goddard98}
\bibinfo{author}{\bibfnamefont{J.~D.} \bibnamefont{Goddard}},
  \emph{\bibinfo{title}{Physics of dry granular media}}
  (\bibinfo{publisher}{Kluwer Academic Publishers}, \bibinfo{year}{1998}),
  chap. \bibinfo{chapter}{Continuum modeling of granular assemblies}.

\bibitem[{\citenamefont{Nedderman}(1992)}]{nedderman92}
\bibinfo{author}{\bibfnamefont{R.~M.} \bibnamefont{Nedderman}},
  \emph{\bibinfo{title}{Statics and kinematics of granular materials}}
  (\bibinfo{publisher}{Cambr. Univ. Press}, \bibinfo{address}{Cambridge},
  \bibinfo{year}{1992}).

\bibitem[{\citenamefont{Frisch et~al.}(1961)\citenamefont{Frisch, Sonnenblick,
  Vyssotsky, and Hammersley}}]{frisch61}
\bibinfo{author}{\bibfnamefont{H.~L.} \bibnamefont{Frisch}},
  \bibinfo{author}{\bibfnamefont{E.}~\bibnamefont{Sonnenblick}},
  \bibinfo{author}{\bibfnamefont{V.~A.} \bibnamefont{Vyssotsky}},
  \bibnamefont{and} \bibinfo{author}{\bibfnamefont{J.~M.}
  \bibnamefont{Hammersley}}, \bibinfo{journal}{Phys. Rev.}
  \textbf{\bibinfo{volume}{124}}, \bibinfo{pages}{1021} (\bibinfo{year}{1961}).

\bibitem[{\citenamefont{Sykes and Essam}(1964)}]{sykes64}
\bibinfo{author}{\bibfnamefont{M.~F.} \bibnamefont{Sykes}} \bibnamefont{and}
  \bibinfo{author}{\bibfnamefont{J.~W.} \bibnamefont{Essam}},
  \bibinfo{journal}{Phys. Rev.}
  \textbf{\bibinfo{volume}{133}}(\bibinfo{number}{9}), \bibinfo{pages}{A310}
  (\bibinfo{year}{1964}).

\bibitem[{\citenamefont{Scher and Zallen}(1970)}]{scher70}
\bibinfo{author}{\bibfnamefont{H.}~\bibnamefont{Scher}} \bibnamefont{and}
  \bibinfo{author}{\bibfnamefont{R.}~\bibnamefont{Zallen}},
  \bibinfo{journal}{The Journal of Chemical Physics}
  \textbf{\bibinfo{volume}{53}}(\bibinfo{number}{9}), \bibinfo{pages}{3759}
  (\bibinfo{year}{1970}).

\bibitem[{\citenamefont{Hicher}(2000)}]{hicher}
\bibinfo{author}{\bibfnamefont{P.-Y.} \bibnamefont{Hicher}}, in
  \emph{\bibinfo{booktitle}{Behaviour of Granular Materials}}, edited by
  \bibinfo{editor}{\bibfnamefont{B.}~\bibnamefont{Cambou}}
  (\bibinfo{publisher}{Springer}, \bibinfo{address}{Wien},
  \bibinfo{year}{2000}), pp. \bibinfo{pages}{1--98}.

\end{thebibliography}
%\makebibliography

\newpage
%\section{Figures}

\end{document}